\pdfoutput=1

\documentclass[11pt]{article}

\usepackage[]{acl}

\usepackage{times}
\usepackage{latexsym}
\usepackage{graphicx}
\usepackage[T1]{fontenc}

\usepackage[utf8]{inputenc}

\usepackage{microtype}
\usepackage{tabularx}

\title{The DialPort tools}

\author{Jessica Huynh\thanks{$^*$Equal contribution} \and Shikib Mehri$^*$ \and Cathy Jiao$^*$ \and Maxine Eskenazi \\ Carnegie Mellon University \\ \texttt{jhuynh, amehri, cljiao, max@cs.cmu.edu} \\}

\begin{document}
\maketitle
\begin{abstract}
The DialPort project (\url{http://dialport.org/}), funded by the National Science Foundation (NSF), covers a group of tools and services that aim at fulfilling the needs of the dialog research community. Over the course of six years, several offerings have been created, including the DialPort Portal and DialCrowd. 
This paper describes these contributions, which will be demoed at SIGDIAL, including implementation, prior studies, corresponding discoveries, and the locations at which the tools will remain freely available to the community going forward.

\end{abstract}

\section{Introduction}
\label{sec:introduction}
The DialPort project\footnote{\url{http://dialport.org/}} has created tools and services that respond to needs voiced by many in the dialog research community during several workshops organized by the Principle Investigators (PIs). Its offerings are available at no cost to the community with the goal of helping researchers gather high quality data, and easily assess and compare their dialog systems. This paper and its corresponding demos showcase the DialPort Portal\footnote{\url{https://dialport.org/portal}} and DialCrowd\footnote{\url{http://dialport.org/dialcrowd.html}}.

There is an increasing need for large amounts of natural dialog data that can be obtained at reasonable cost and in an interactive manner. Static datasets are ineffective for both evaluation and optimization. 
This has led to the creation of the DialPort Portal, which facilitates the collection of flexible and evolving data as well as interactive assessment with real users. Notably, the Portal was used to connect systems and collect data for the \textit{Interactive Evaluation of Dialog} track \citep{mehri2021interactive} at DSTC9 \citep{gunasekara2020overview}.

Another community need centers around how to gather high quality data when using crowdsourcing platforms. DialCrowd has been constructed to facilitate crowdsourcing by guiding researchers to give clear, understandable explanations of the task to the workers who produce or annotate data. It also aids in calculating the correct level of worker payment. Finally, it includes several methods of data quality assessment.

The University of Southern California (USC) is a partner in DialPort. The team at USC works on a tools repository\footnote{\url{https://dialport.ict.usc.edu/}} and the REAL Challenge. 

This paper gives background and describes in detail the parts of both the Portal and DialCrowd. It also provides information on how to access and use them. As the DialPort project draws to an end, the paper indicates the permanent sites where these tools will reside.

\section{Background}
\label{sec:background}

\subsection{Interactive Platforms for Dialog}

As dialog models improve, it is imperative that they are evaluated in interactive settings with real users. \citet{mehri2020unsupervised} show that while pre-trained dialog systems excel at generating responses \citep{zhang2019dialogpt,bao2020plato}, they underperform in back-and-forth interactions.

The Alexa Prize challenge \citep{ram2018conversational,khatri2018advancing} allows university teams to build socialbots that are assessed in interactive settings with Alexa users. In contrast, the DialPort Portal is accessible to the broader research community. Furthermore, the Alexa Prize challenge primarily relies on speech input from the user, which may result in speech recognition errors. Though the DialPort Portal can accept speech input, its web interface can also be used with text-only input.

\subsection{Crowdsourcing}

With the amount of dialog data available or able to be collected with systems such as DialPort, it is important to have easy and accessible tools to create detailed annotations of this data for different metrics. One method of obtaining annotations is crowdsourcing with platforms such as Amazon Mechanical Turk (AMT). However, it is sometimes difficult to obtain conclusive results, and a survey of current natural language processing HITs has shown the weaknesses of these HITs \cite{huynh2021survey}. Instructions \cite{chandler2013risks}, examples \cite{doroudi2016toward}, and payment are some of the aspects that need to be attended to in order for HITs to acquire higher quality data.

\section{DialPort Portal}
\label{sec:portal}

The DialPort Portal was initially conceived with the objective of listing many dialog systems from a variety of sites. This type of platform, with demonstrations, links, and references to various systems, is valuable to both researchers and real users. The concept of the Portal evolved, and the different systems were linked such that a user could interact with all of the connected systems, transitioning seamlessly between systems, with the dialog state (consisting of slots such as city or date) shared across systems \citep{zhao2016dialport,lee2017dialport}. As dialog systems continued to improve, especially with the advent of engaging response generation models \citep{zhang2019dialogpt,bao2020plato}, the Portal recruited real users through Facebook advertising with the objective of providing researchers with a platform to collect interactive dialogs with real users \citep{mehri2021interactive}. 

\subsection{Portal Version 1}

The original version of the Portal grouped several dialog systems from different sites (Cambridge, USC, CMU) and managed seamless switching amongst \citep{zhao2016dialport}. For example, a user could ask for the weather in Pittsburgh and get the CMU weather system, then ask the CMU system for the weather in Cambridge, then ask for a restaurant and automatically switch to the Cambridge restaurant system, then ask to play a game and get the USC system.

This instance of the Portal serves as a platform to interact with different systems over the course of one dialog \citep{zhao2016dialport}. To accomplish this, the Portal needed to address several challenges (1) how to share information across systems (e.g., remembering the city the user wanted the weather for when interacting with the CMU system, and sharing that with the Cambridge system when the user wants a restaurant recommendation), (2) how to gracefully continue a dialog when a system is down, and (3) how to give two systems addressing the same task (e.g., restaurants) equal time with the users. Respectively, these problems were addressed by (1) maintaining a shared dialog state across systems, (2) backing off to an equivalent system or changing the topic, and (3) a pseudo-random system selection policy. In order to make the system easy to use, an API was developed to facilitate connecting new systems to the Portal. This version has pedagogical value as it can easily be demonstrated for dialog classes.

\subsection{Portal Version 2}

\begin{figure*}
    \centering
    \includegraphics[width=\textwidth]{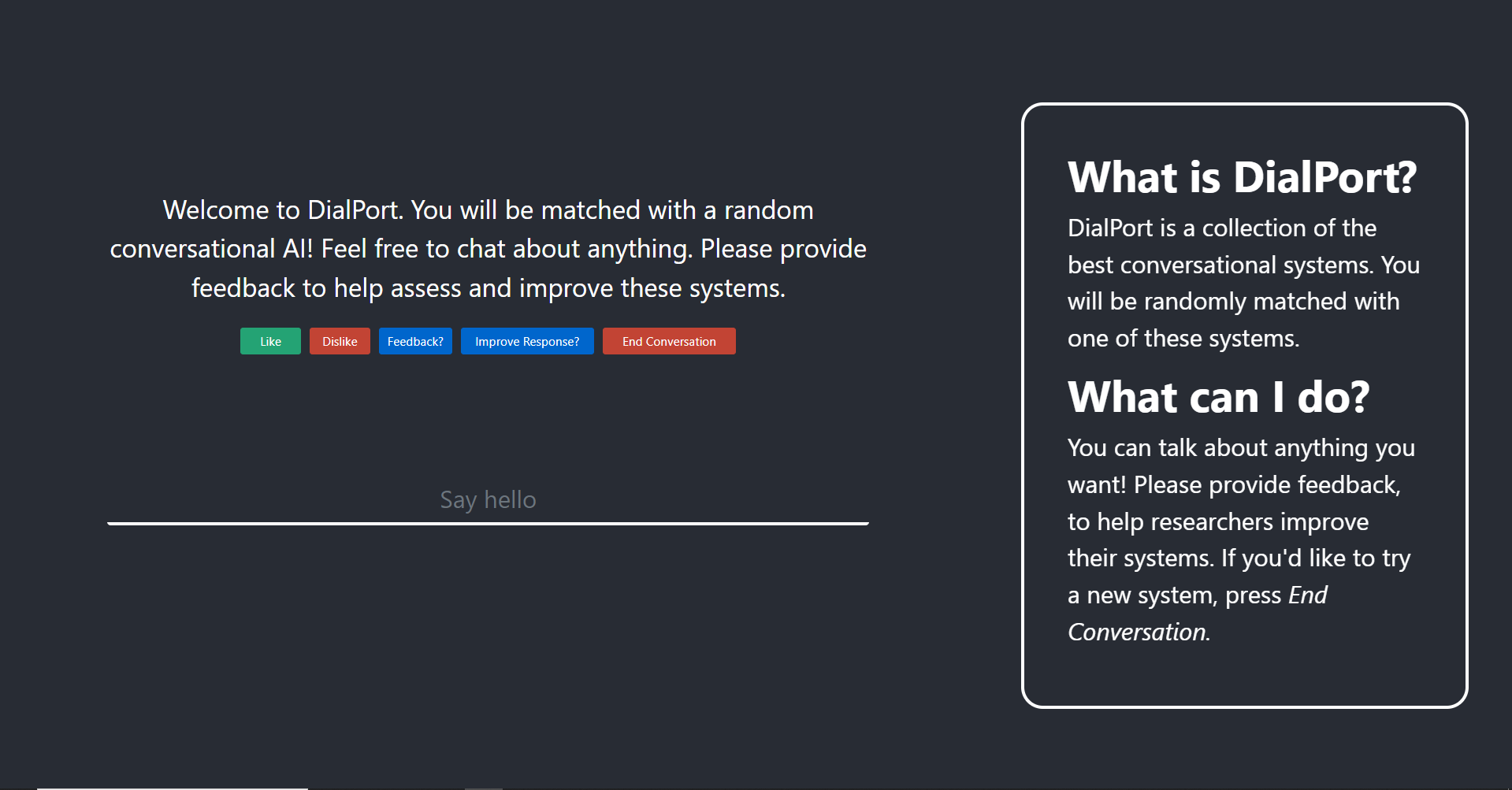}
    \caption{DialPort Portal. This screenshot of the Portal displays (1) the dialog history, shown in the center of the screen, (2) an input field for the user to type their responses, and (3) a set of feedback buttons below the dialog history (``Like'', ``Dislike'', ``Feedback?'', ``Improve Response?'' and ``End Conversation''). The interface clear and emphasizes the three important actions that a user should perform while using the Portal: (1) reading the dialog history, (2) responding to the dialog system, and (3) providing feedback.}
    \label{fig:my_label}
\end{figure*}

With the advent of the API, the possibilities of use of the Portal greatly expanded. The Portal was used for the DSTC9 Challenge \citep{mehri2021interactive}, as a tool that enabled researchers to both compare their systems on one common platform (with real users) and to gather considerable amounts of data. The Portal was made available to DSTC9 participants. The idea was to connect systems and have them tested by real (unpaid) users. The CMU DialPort team advertised the Portal on Facebook and interested individuals tried it out (with text only). Upon visiting the Portal, real users are randomly matched with a dialog system, without knowledge of the specific system they are interacting with. While some people left the site after only one or two turns with a system, many actually continued to communicate with a system for a substantial conversation, and were thus considered to be real users. Real users consist of users who find some personal interest (getting information, companionship, curiosity) in continuing a dialog. There were 11 participants in the interactive part of the Challenge \citep{mehri2021interactive}. With an advertising budget of \$2500, we collect more than 4000 dialogs on the DialPort portal (2960 dialogs with at least 4 turns or 8 utterances); thus the cost was less than \$1.00 per usable dialog. The DialPort portal, through funding from the National Science Foundation, has been able to provide interactive evaluation as a service free of charge to any dialog researchers. The Appendix contains a sample dialog from the winning system of the DSTC9 track \citep{bao2020plato}.

DSTC9 demonstrated that the Portal could easily be used to both compare systems and to gather data with real users. Besides challenges, another potential use of the Portal would be for students to connect systems that they build for a class project to see how well they do in real user interaction.

At the end of the DialPort project in the coming year, the Portal will move from the Dialog Research Center at CMU to LDC at UPenn.

\subsection{DialPort Dashboard}
After collecting data from real users on the DialPort interface, a subsequent task is to perform analysis on the gathered data. We provide the DialPort dashboard which allows researchers to (i) \textit{analyze} dialogs collected on their system, (ii) \textit{interact} with the dashboard to filter and organize dialogs based on various criteria, and (iii) \textit{compare} their system to other systems connected to the DialPort Portal. Currently, the Dashboard contains over 7000 dialogs from 28 systems. The Dashboard is connected to the DialPort Portal via API calls, allowing dialogs to be quickly displayed on the Dashboard after being collected from the Portal. The Dashboard code will soon be released, allowing for use of the Dashboard in offline mode. 

\label{sec:dialcrowd}

The Dashboard UI contains panels, tables, and charts. At both the system and dialog level, attributes such as the number of utterances, likes, dislikes, comments, corrections are displayed (see figure \ref{fig:dashboard-main}). In addition, the two evaluation metrics of FED \citep{mehri2020unsupervised} and human ratings are shown. Since the Dashboard is designed to be easily extended, additional metrics can be added in the future. Users can interact with the dashboard by filtering and ranking dialogs based on attributes and metrics. For example, the provided toolbar can be used to find all conversations with a given user's system with more than \textit{n} turns or rank conversations from most-to-least number of likes. Users can also filter words and phrases in dialogs by their number of occurrences from the perspective of both the system or human participant, and thus view common phrases or words mentioned on either side of the conversation. Finally, each system contains a progress monitor graph which displays the number of dialogs being collected over time, allowing users to actively observe data collection in the DialPort Portal.

\begin{figure*}[h]
\centering
\includegraphics[width=\textwidth]{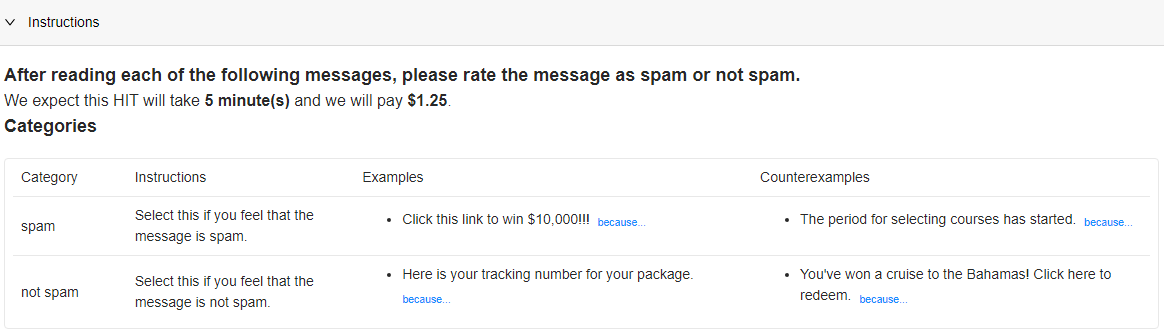}
\caption{DialCrowd Examples and Counterexamples with Explanations}
\label{fig:dialcrowd}
\end{figure*}

\section{DialCrowd}

To address the many issues that present themselves when using crowdsourcing to collect high quality data, DialCrowd was created. DialCrowd \cite{lee2018dialcrowd} is a dialog assessment toolkit which aids researchers with human intelligence task (HIT) creation. Requesters follow templates on the DialCrowd site, which generate a HIT that can be linked for a worker on any crowdsourcing site.

The second version of this tool \cite{huynh-EtAl:2022:LREC} focuses on collecting high-quality data with tools such as:
\begin{itemize}
    \setlength{\itemsep}{0pt plus 1pt}
    \item Links to create better instructions
    \item Prompts to provide examples and counterexamples with explanations seen in Figure \ref{fig:dialcrowd}
    \item Functionality for adding golden data and duplicate data in each HIT
    \item Payment suggestions
    \item A feedback area
    \item Overall statistics from the HIT (time, patterns in the responses, inter-annotator agreement)
\end{itemize}
This allows for requesters to create a well-structured HIT which allows workers to provide better quality annotations. Consequently, it makes it easier to filter responses from potential bots. Additional tools include the capability to include a mandatory consent form at the start of the HIT, and detailed style changes for the HIT. Further description of the system along with corresponding images can be found in \cite{huynh-EtAl:2022:LREC}.

One DialCrowd template, intent classification, has been merged into the new home for DialCrowd, ParlAI \footnote{\url{https://github.com/facebookresearch/ParlAI/tree/main/parlai/crowdsourcing/tasks/dialcrowd}}, and is now available for use.

\section{The DialPort demo}
\label{sec:dialportdemo}
The demos of the DialPort Portal and Dashboard and of DialCrowd at SIGDIAL will include:
\begin{itemize}
    \setlength{\itemsep}{0pt plus 1pt}
    \item how to connect a system
    \item what interaction with each tool looks like
    \item advantages there are in using the tools, with examples (for example, what resulting data looks like)
\end{itemize}

\section{Conclusion and Future Directions}
The tools presented in this demo help dialog researchers in data gathering and assessment. As the community uses them, more types of applications will arise. The tools have been created in a way that enable additions as the field and the needs evolve.

\section{Acknowledgements}
\label{sec:acknowledgements}

This work is funded by National Science Foundation Grant Nos. CNS-1512973, DGE1745016, and DGE2140739. The opinions
expressed in this paper do not necessarily reflect
those of the National Science Foundation. The authors would like to thank Tiancheng Zhao, Kyusong Lee, and Ting-Rui Chiang for their contributions to these tools.

\bibliography{anthology,custom}
\bibliographystyle{acl_natbib}

\appendix

\vspace{1mm}

\section{Appendix}
\label{sec:appendix}

\begin{table}[ht]
\renewcommand*{\arraystretch}{1.2}

    \centering
    \begin{tabular}{|m{\linewidth}|}
    \hline
\textbf{User:} Hello\\
\textbf{System:} hello! hows your day going so far?\\
\textbf{User:} What are you up to?\\
\textbf{System:} just finished making coffee! what about you? what do you do?\\
\textbf{User:} i'm playing video games\\
\textbf{System:} oh, nice! what kind? and i'm just relaxing here and listening to some classical music.\\
\textbf{User:} i'm playing league of legends\\
\textbf{System:} nice! what rank are you? i haven't played in forever.\\
\textbf{User:} who is your favorite classical musician?\\
\textbf{System:} probably beethoven. i like classical music.\\
\textbf{User:} i'm plat 2\\
\textbf{System:} nice! i'd imagine you are better than that!\\
\textbf{User:} can you play an instrument?\\
\textbf{System:} i can play the piano. do you play any instruments?\\
\textbf{User:} thanks\\
\textbf{System:} i appreciate it! so what is it like being plat 2?\\
\textbf{User:} i can play the clarinet, but not very well haha\\
\textbf{System:} oh nice! i played clarinet in high school.\\
\textbf{User:} the games are difficult but rewarding\\
\textbf{System:} yes they are! i played the trumpet in college.\\
\multicolumn{1}{|c|}{.   .   .} \\ \hline
        
    \end{tabular}
    \caption{An example of a conversation snippet from the DSTC9 top performing system. The system is capable of maintaining two simultaneous threads of dialog about League of Legends and music, and switch between them in a coherent and engaging manner.}
    \label{tab:example_dialog2}
\end{table}

\begin{figure*}
    \centering
    \includegraphics[width=0.9\textwidth]{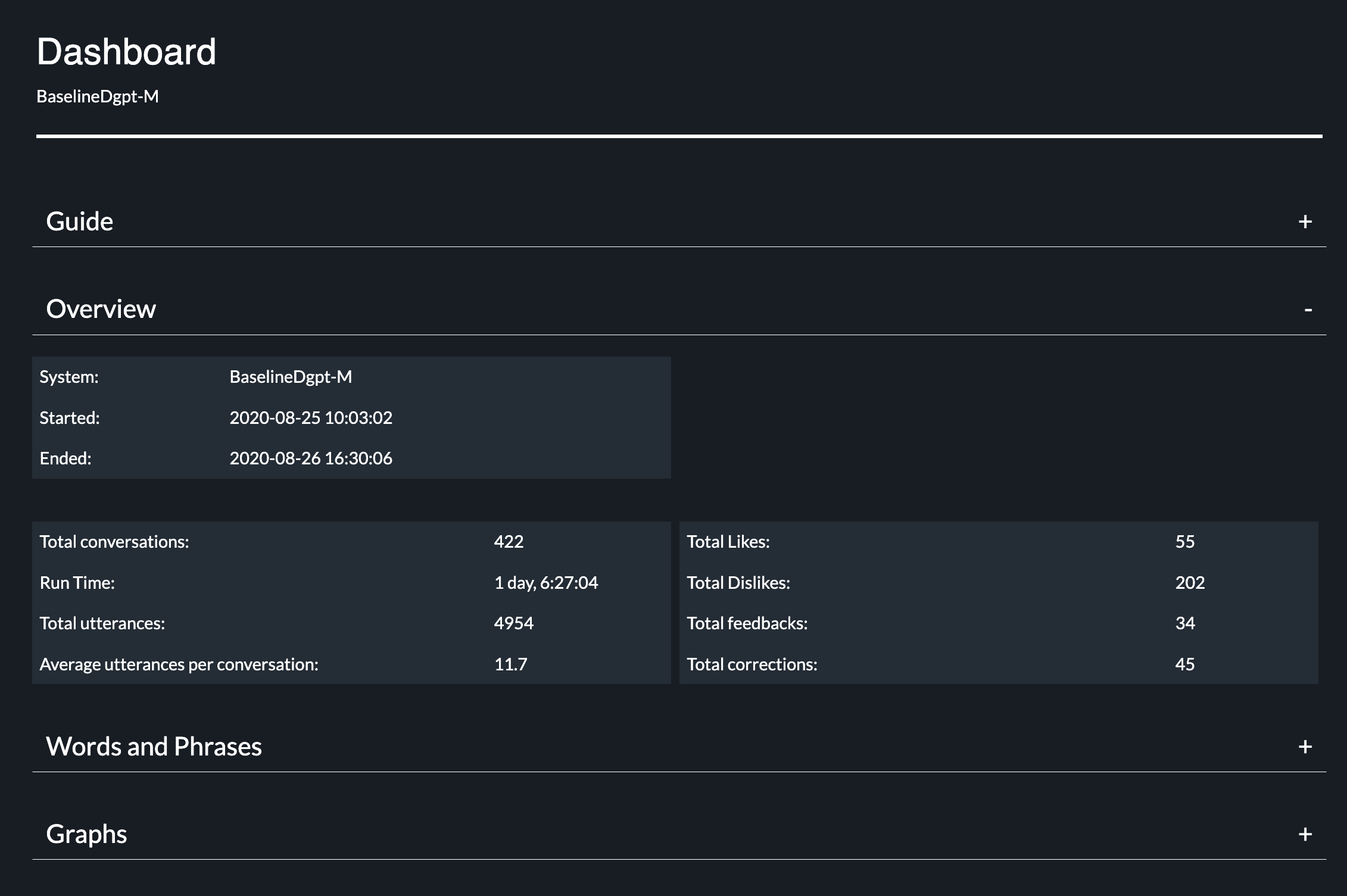}
    \caption{The home page for a system on the DialPort dashboard. General information about the conversations collected from the system are displayed. Sections such as "Words and Phrases" and "Graphs" can be expanded or collapsed to view additional information about the system.}
    \label{fig:dashboard-main}
\end{figure*}

\begin{figure*}
    \centering
    \includegraphics[width=0.75\textwidth]{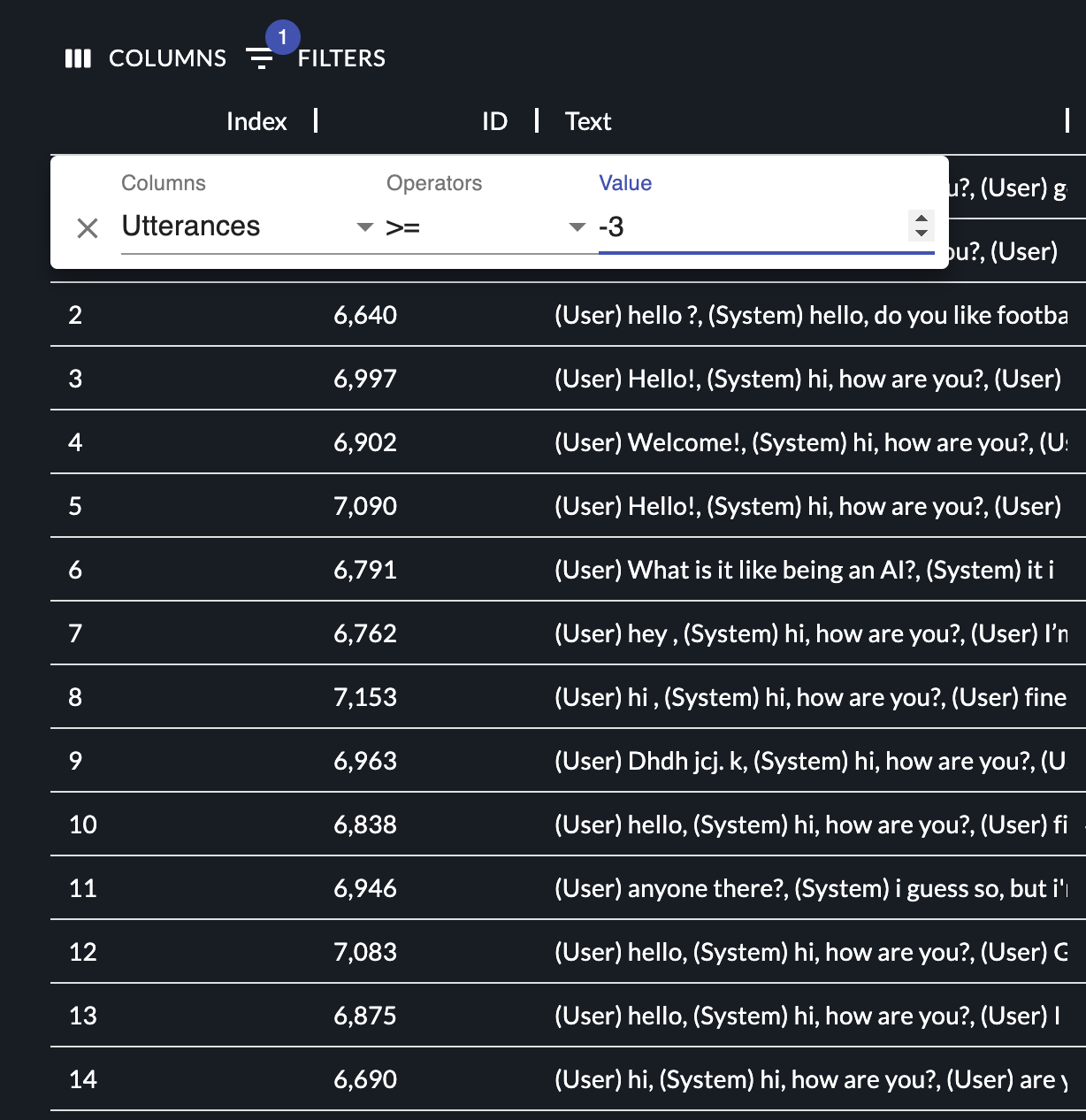}
    \caption{Using the DialPort dashboard to find all conversations in a system with more than 3 utterances}
    \label{fig:dashboard-filter}
\end{figure*}

\end{document}